\documentclass[12pt]{article}
\usepackage{amssymb}
\usepackage{amsfonts}
\usepackage{amsmath,amsthm}
\usepackage{graphics,epsfig}
\usepackage{subfigure}
\usepackage{graphicx,epsfig, color }

\oddsidemargin 10mm \numberwithin{equation}{section}
\def\be{\begin{equation}}
\def\ee{\end{equation}}
\begin{document}
\begin{center} {{\bf {Effects of a cloud of strings on the extended phase space of Einstein-Gauss-Bonnet AdS black holes}}\\
 \vskip 0.50 cm
  {{ Hossein Ghaffarnejad \footnote{E-mail: hghafarnejad@semnan.ac.ir
 } }{and Emad Yaraie \footnote{E-mail: eyaraie@semnan.ac.ir
 } }}\vskip 0.2 cm \textit{Faculty of Physics, Semnan
University, P.C. 35131-19111, Semnan, Iran }}
\end{center}
\begin{abstract}
In this paper we study the thermodynamics of Einstein-Gauss-Bonnet
(EGB)-AdS black holes minimally coupled to a cloud of strings in
an extended phase space where the cosmological constant is treated
as pressure of the black holes and its
 conjugate variable is the thermodynamical volume of the black holes. To investigate the analogy between EGB black holes
  surrounded by a cloud of strings and liquid-gas system we derive the analytical solutions of the critical points and probe the effects of a
  cloud of strings on $P-V$ criticality. There is obtained resemblance between "small black hole/large black hole" (SBH/LBH) phase transition and
   the liquid-gas phase transition.
   We see that impact of a cloud of strings can bring Van
   der Waals-like behavior, in absence of the Gauss-Bonnet (GB) counterpart. In the other words, in the EGB black hole with $\alpha \to 0$
   and when it is surrounded by a
    cloud of strings the Hawking-Page phase transition would be disappeared and SBH/LBH phase transition
    recovers. Also there is not happened Joule-Thomson effect.
\end{abstract}
\section{Introduction}
Black holes are one of the fascinating predictions of the
Einstein's theory of general relativity. The study of black hole
as a thermodynamical system have been an attractive subject in
theoretical physics for many years. In the recent years, the study
of the thermodynamic properties of black holes have revealed many
aspects of them. One of these aspects is thermodynamic phase
transition in AdS black holes [1].
 First works about the AdS black
holes thermodynamics
 obtained from higher derivative gravity was studied in ref. [2]
 and for EGB type one can follow [3].  The study of the AdS black hole phase transition has
 been generalized to the extended phase space where the cosmological constant has been treated as the pressure of the black hole [4,5].
 As we can see in [6] Van der Waals-like behavior of Reissner-Nordstrom black hole could be observed. For more study one can see [7-22].
  The authors in [23] constructed a novel 5-dimensional black hole solution in the
EGB gravity and in a cloud of strings background, then studied its
aspects in a non-extended thermodynamics.
 Because the universe can be described by one-dimensional objects called strings, this solution has a great physical
 significance:
 A cloud of strings is introduced as a configuration of one-dimensional strings from the string
  theory as the most promising theory of quantum gravity which could be effective on gravitational fields such as black holes. This extension
   has some advantages like the ability of studying any higher dimensions and resemblance energetic field such as the global monopole in
   four dimension. Since strings are supposed to be fundamental objects in nature which are supported by observations, so it encourages us to study its
   effects on various gravitational theory. So the importance of studying the gravitational effects of the matter which is in the form of a cloud of
    strings arises. At first Strominger and Vafa showed a connection between counting string states and the entropy of the black hole in [24]
    , later in
    [25] it was considered a model for a cloud of strings that would be the equivalent of a perfect fluids. Implications of that for a broad range of
    black holes have been studied by many authors [23-32], as we can see some earlier works performed these effects on gravitational theories like
    Schwarzschild solution in [50], higher derivative theories like Lovelock theory in [30, 31, 32] and modified gravity in [51]
    , and also by
     considering the effect of quintessence dark energy in [52, 53].
     Gauss Bonnet terms in higher derivative theories could be seen in the low-energy effective action of superstring theories.
     They could be viewed as the corrections of large N expansion of boundary in the context of
      AdS/CFT duality in the strong coupling limit. Since such corrections have interesting effects, so it would be natural to study this black hole
      solution surrounded by a spherically symmetric string clouds as a thermodynamic system and seek criticality behavior and the effect of coupling
      constant of the model like the case for $\alpha_{GB}\rightarrow0$ which leads to Schwarzschild black hole. We can also seek how the effect of the
       background string clouds can alter phase transition of black hole.
In fact, this work is an extension of the previous work [10] to a
cloud of strings background where we study thermodynamic aspects
of the black hole with extended
 phase space and investigate the effects of a cloud of strings on $P-V$ criticality. We also calculate critical exponents of the system.\\
 Layout of the paper is as
 follows.
 In section 2 we define higher EGB
higher derivative gravity in presence of a cloud of strings
 effects. We calculate equation of first law EGB black hole
 thermodynamics. In section 3 we use extended phase space
 thermodynamics to obtain its critical points at P-V hypersurface.
 Critical exponents are calculated in section 4 and finally we seek possibility of creation of the Joule Thomson
 effect. Section 5 denotes to conclusion and outlook of this work.
\section{EGB black holes  with  cloud of strings}
Let us we start with action of EGB gravity surrounded by a cloud
of strings which in 5- dimension is given by [23]
\begin{equation}
 S = \frac{1}{2}\int \sqrt{-|g|} \big(R - 2 \Lambda + \alpha \mathcal{L}_{\mathrm{GB}}\big) d^{5}x
         + I_{\mathrm{NG}},
\end{equation}
where $\alpha>0$ is the GB coupling coefficient, $|g|$ is absolute
value of determinant of the metric field $g_{\mu\nu}$ and $
 \mathcal{L}_{\mathrm{GB}} = R_{\mu \nu \rho \sigma} R^{\mu \nu \rho \sigma} + R^{2} - 4 R_{\mu \nu} R^{\mu \nu},
$ is the GB counterpart of the lagrangian which originates from
quantum fields renormalization. Dynamics of a classical
relativistic string is described by the Nambu-Goto action
$I_{\mathrm{NG}} = \int_{\Sigma} p \sqrt{-|\gamma |} d\lambda^{0}
d\lambda^{1}$, where $(\lambda^{0}, \lambda^{1})$ are local
coordinates of the string
 which makes parameterized the worldsheet. $|\gamma|$  is absolute value of determinant of an induced metric $\gamma_{a b}$ on the strings
 worldsheet for which
 $\gamma_{a b} = g_{\mu \nu} \frac{\partial x^{\mu}}{\partial \lambda^{a}} \frac{\partial x^{\nu}}{\partial
 \lambda^{b}}$.
On the other hand the bivector related to the strings worldsheet
can be written as $ \Sigma^{\mu \nu}=\epsilon^{ab} \frac{\partial
x^{\mu}}{\partial \lambda^{a}} \frac{\partial x^{\nu}}{\partial
\lambda^{b}}, $ where $\epsilon^{ab}$ is Levi-Civita tensor. So
the energy momentum tensor for a cloud of strings is given by $
T^{\mu \nu}=(-\gamma)^{-\frac{1}{2}}\rho \Sigma^{\mu
\sigma}\Sigma_{\sigma}^{\nu}, $ in which the proper density of a
string cloud is described by $\rho$. Applying (2.1) and the static
spherically symmetric metric
$ds^2=f(r)dt^2-f(r)^{-1}dr^2-r^{2}d^{2}\Omega$
 one can obtain metric field equation as follows.
\begin{equation}
\frac{1}{2 r^{3}} \Big(2 r^{3} \Lambda + 6 r f(r) - 6 r + 3 r^{2}
f(r)'\Big)+ \frac{6}{ r^{3}}\alpha f(r)' (1 - f(r)=-\frac{a}{r^3},
\end{equation}
\begin{equation}
 -1 + r^{2} \big(\Lambda + \frac{f(r)''}{2}\big) + f(r) +2rf(r)'-2 \alpha f(r)'^{2}+ 2 \alpha f(r)'' (1 - f(r))=0,
\end{equation}
with solution
\begin{equation}
     f(r) = 1 + \frac{r^{2}}{4 \alpha} \left(1 - \sqrt{1+\frac{32\alpha M}{r^{4}} -\frac{8\alpha}{\ell^2}
            + \frac{16 a \alpha}{3 r^{3}}} \right),
\end{equation}
 where $a$ is real positive constant and $\ell$ corresponds to the AdS radius. It is related to the pressure of the black hole through
 $P=-\frac{\Lambda}{8\pi}=\frac{3}{4\pi\ell^2}$ in 5 dimension.
 Location of the event horizon is obtained by solving $f(r_{+})=0$, so the ADM mass M, entropy S and
 Hawking temperature of the black hole can be derived respectively
 as follows.
\begin{equation}
M=\frac{1}{3}P\pi
r_{+}^4-\frac{1}{6}ar_{+}+\frac{1}{4}r_{+}^2+\frac{\alpha}{2},
\end{equation}
\begin{equation}
S=\int^{r_+}_0\frac{1}{T}\left(\frac{\partial M}{\partial
r_+}\right)dr_+=\pi(\frac{r_+^3}{3}+4\alpha r_+),
\end{equation}
\begin{equation}
T=\frac{f'(r_+)}{4\pi}=\frac{1}{6\pi}\frac{8P\pi
r_{+}^3-a+3r_{+}}{r_{+}^2+4\alpha}.
\end{equation}
It is interesting to note that the entropy of a cloud of strings
does not affect on the black hole entropy. The black hole mass $M$
in the extended thermodynamics is treated as enthalpy, so with
respect to the above thermodynamics definitions the first law of
the black hole thermodynamics in an extended phase space reads
\begin{equation}
dM=TdS+VdP+\mathcal {A}da+\mathcal{B}d\alpha,
\end{equation}%
where $V=\left(\frac{\partial M}{\partial
P}\right)_{S,a,\alpha}=\frac{1}{3}\pi r_+^4$ is the thermodynamic
volume, $\mathcal {A}=\left(\frac{\partial M}{\partial
\alpha}\right)_{S,P,a}=\frac{1}{2},$ and $\mathcal
{B}=\left(\frac{\partial M}{\partial
a}\right)_{S,P,\alpha}=-\frac{1}{6}r_{+},$ stand for the physical
quantities conjugated to the parameters $a$ and $\alpha$
respectively. Physical meanings of the conjugated potentials
$\mathcal {A}$ and $\mathcal {B}$ needs further investigation.
Using the scaling argument we can obtain the generalized Smarr
relation for the GB AdS black hole in the presence of cloud of
strings as follows.
\begin{equation}
2M=3TS-2VP+2\mathcal{A}\alpha+\mathcal{B}a.
\end{equation}
Finally we would like to discuss some about the constraints and
singularities: by attention to
 ansatz (2.4) in order to have a well defined (without any imaginary or naked singularity [45]) vacuum solution ($M=0$ and $a=0$), GB
 coefficient $\alpha_{GB}$ must be restricted as $0\leq\alpha_{GB}\leq\ell^2/8$. Also for holographic purposes $\alpha_{GB}$ is also
 constrained as $-7/72\leq\alpha_{GB}\leq9/200$ due to causality and positive definiteness of the boundary energy density [46, 47].
 The non-vacuum solutions $f(r)$ must be also a real-valued function, so in general we encounter by two types of singularities: Causal
 singularity at $r=0$ and branch  singularity for $r>r_{br}$ in which $r_{br}$ is defined
 when the square root in (2.4) is non-negative leads to minimum non-negative real root of the following polynomial equation
\begin{equation}
r_{br}^4\big(1-\frac{8\alpha}{\ell^2}\big)+16\alpha\big(\frac{ar_{br}}{3}+2M\big)=0.
\end{equation}
Therefore the allowed domain of radius from $0<r<\infty$ reduces
to $0<r<r_{br}$. On the other side if $r\rightarrow0$ then $f(r)$
 approaches $1-\sqrt{\frac{m}{\alpha}}$ which demonstrates that usual singularity at $r=0$ removed by GB term.\\
It is interesting to discuss some about the causality constraints
in our model which comes from the requirement that the equation
 of motions of perturbation must be hyperbolic [48,49], so the perturbations propagate in a causal way. The hyperbolicity
 condition of the equation of motion of Lovelock theories as necessary condition for causality is equivalent to Lorentzian effective
 metric in field space. So if the effective metric be non-Lorentzian, it implies causality violation. The effective metric components
  for 5-dimensional GB theory regarding [49] are:
\begin{equation}
[G^{\alpha\alpha}]^{\tilde{i}}_{\tilde{j}}=g^{\alpha\alpha}\bigg(1-2\alpha\frac{f'(r)}{r}\bigg),~~~
[G^{kk}]^{\tilde{i}}_{\tilde{j}}=g^{kk}\bigg(1-2\alpha
f''(r)\bigg),
\end{equation}
in which $g^{\alpha\alpha}$ and $g^{kk}$ are Lorentzian, so the
signature of effective metric depends on factors. Large values of
$r$ lead to a Schwarzschild behavior of the metric and so the
effective metric would be Lorentzian. But the case for small $r$
is different: it can be seen that for all values of $M$,
$\alpha_{GB}$, $\ell$ and $a$ the factor in
$[G^{kk}]^{\tilde{i}}_{\tilde{j}}$ is negative, $1-2\alpha
f''(r)<0$, but factor in
$[G^{\alpha\alpha}]^{\tilde{i}}_{\tilde{j}}$ can be positive or
negative. Actually by attention to (2.4) it is simple to find $
r_*\simeq3.36\big(\frac{a\alpha}{\frac{8\alpha}{\ell^2}-1}\big), $
that when $r>r_*$ factor
$[G^{\alpha\alpha}]^{\tilde{i}}_{\tilde{j}}$ would be negative and
therefore the effective metric takes Lorentzian signature, however
for $r<r_*$
 causality violations happens.
\section{Thermodynamics behavior }
Applying (2.7) and $V=\left(\frac{\partial M}{\partial
P}\right)_{S,a,\alpha}=\frac{1}{3}\pi r_+^4$ one can obtain
equation of state of this black hole as follows.
\begin{equation}
P=\frac{T(v^2+\frac{64}{9}\alpha)\pi+\frac{8}{27}a-\frac{2}{3}v}{\pi
v^3}
\end{equation}
where $v=\frac{4}{3}r_{+}$ is called as the specific volume [4].
Solving $\left.\frac{\partial P}{\partial v}\right|_{T=T_c}=0$ and
$\left.\frac{\partial ^2P}{\partial v^2}\right|_{T=T_c}=0,$  one
can obtain the critical points as follows.
\begin{equation}
v_c=\frac{2(a+\sigma)}{3},~~T_c=\frac{\sigma}{\pi(a^2+a\sigma+48\alpha)},~~P_c=\frac{a^3+a^2\sigma+48a\alpha+24\sigma
\alpha}{\pi(a^2+a\sigma+48\alpha)(a+\sigma)^3},
\end{equation}
where $\sigma=\sqrt{a^2+48\alpha}$ the subscript $c$ denotes to
the critical point. By comparing our results with [10] for
uncharged case it would be interesting to note that the effect of
a cloud
 of strings reflected in all critical points which by putting $a=0$ these critical points reduced to the result of an uncharged
  GB black hole solution. Note that universal Van der Waals ratio in this case depends on the values of GB and string cloud factors.
We plot the $P-v$ isotherm graph with different values of string
cloud parameter $a=\{0.1,1\}$ and for weak GB counterpart
$\alpha=0.001$ in figure 1-a. The solid lines correspond to the
ideal gas phase transition when temperature is above the critical
value which shows a monotonically decreasing behavior. Upper than
the critical temperature the dashed lines can be divided to three
different branches. They indicate to large, small and medium black
hole branches. The medium black hole branch is unstable while the
large black hole and the small black hole branches are stable
which mimic the Van der Waals liquid/gas phase transition. In
figure 1-b we plotted $P-v$ diagrams for temperature below the
critical temperature in which $\alpha$ alters. We can see  the
diagram takes a shifting and pressure goes to be negative by
increasing GB coefficient. Also there exists a particular
temperature $T_{0}$ similar to the one which is obtained for Van
der Waals fluid. It is achieved by solving $\frac{\partial
P}{\partial v}=P=0,$ which  below this temperature the pressure is
not be positive for some horizon radius, and of course could be
remedied by Maxwell construction. However $\frac{\partial
P}{\partial v}=P=0,$ reads $v_{0}=\frac{4}{9}(a+\xi )$ and
$T_{0}=\frac{3}{4}\frac{2\xi -a}{\pi(a\xi +a^2+72\alpha)}$, where
$\xi=\sqrt{a^2+36\alpha}$ and is depicted for $a=1$ and
$\alpha=0.01$ in figure 1-c. It is of vital importance to study
the behavior of Gibbs free energy to analysis global stability of
the system. In the extended phase space the Gibbs free energy is
given by $G=M-TS$  and for the black hole under consideration
takes the form
\begin{equation}
G=-\frac{1}{36}\frac{4P\pi r^6+144P\pi \alpha
r^4+4ar^3-3r^4+18\alpha r^2-72 \alpha^2}{r^2+4\alpha}.
\end{equation}
\begin{figure}[tbp] \centering
    \subfigure[{}]{\label{1}
    \includegraphics[width=.31\textwidth]{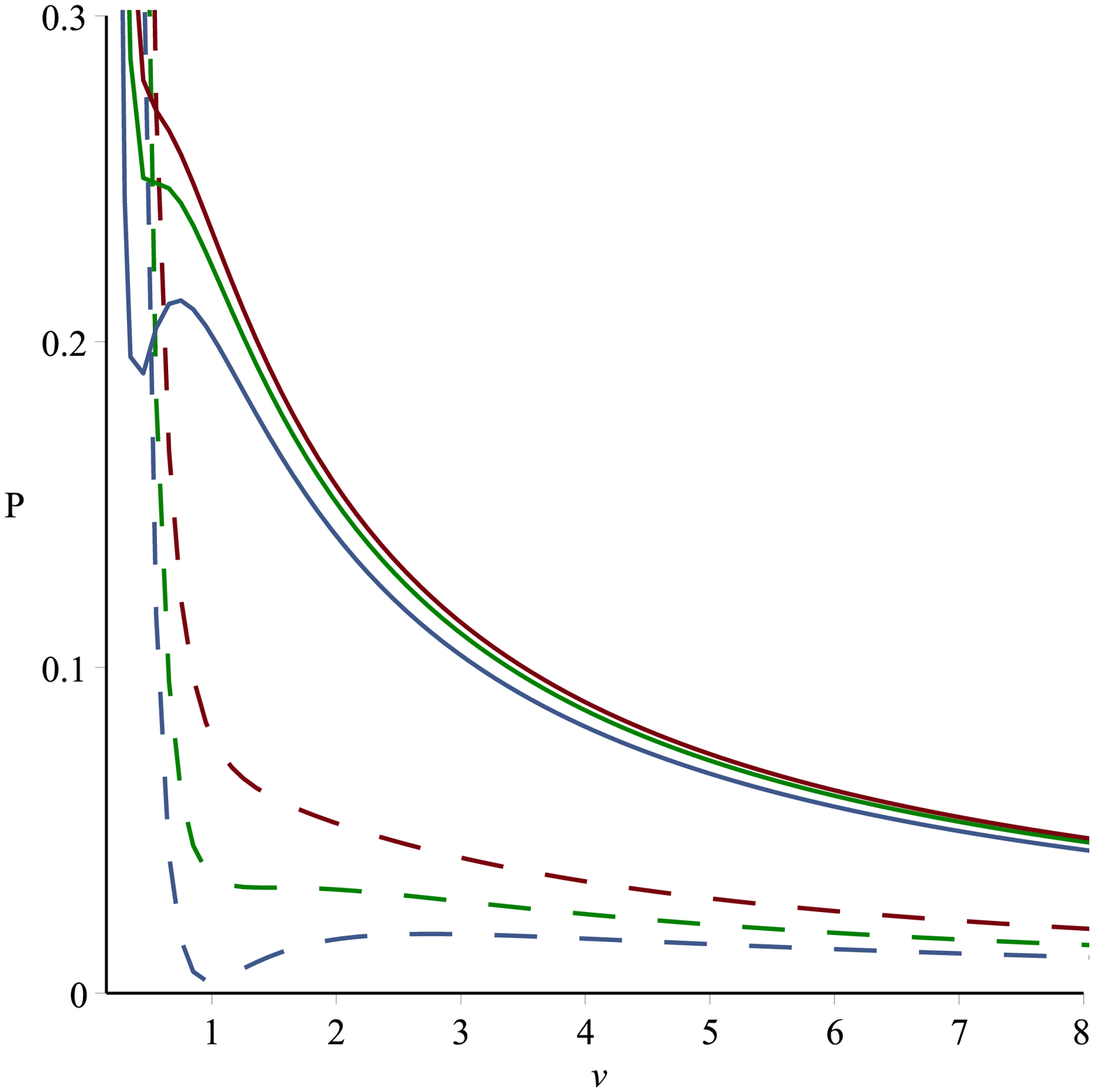}}
\hspace{1mm} \subfigure[{}]{\label{1}
\includegraphics[width=.31\textwidth]{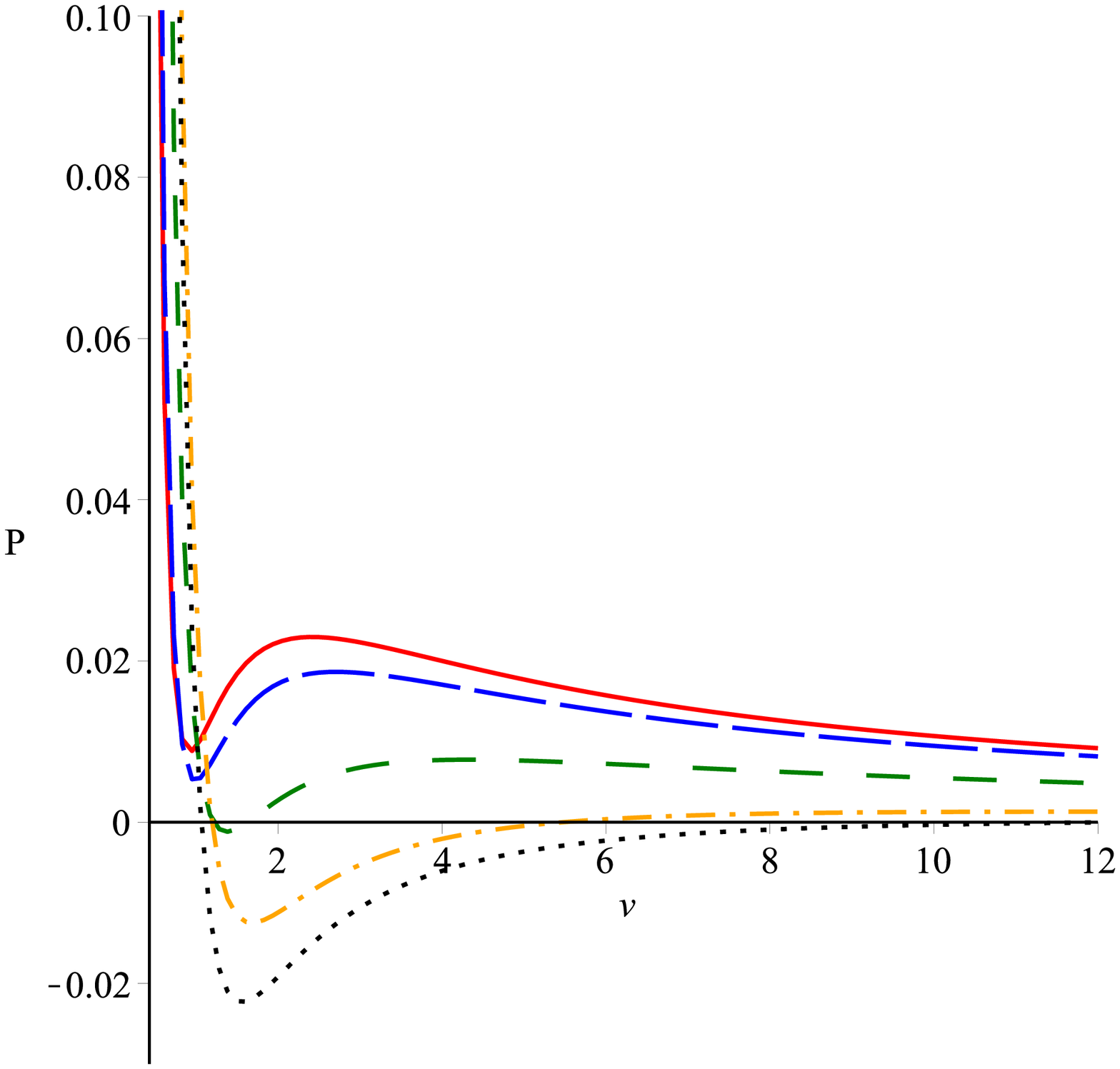}}
   \subfigure[{}]{\label{1}
    \includegraphics[width=.31\textwidth]{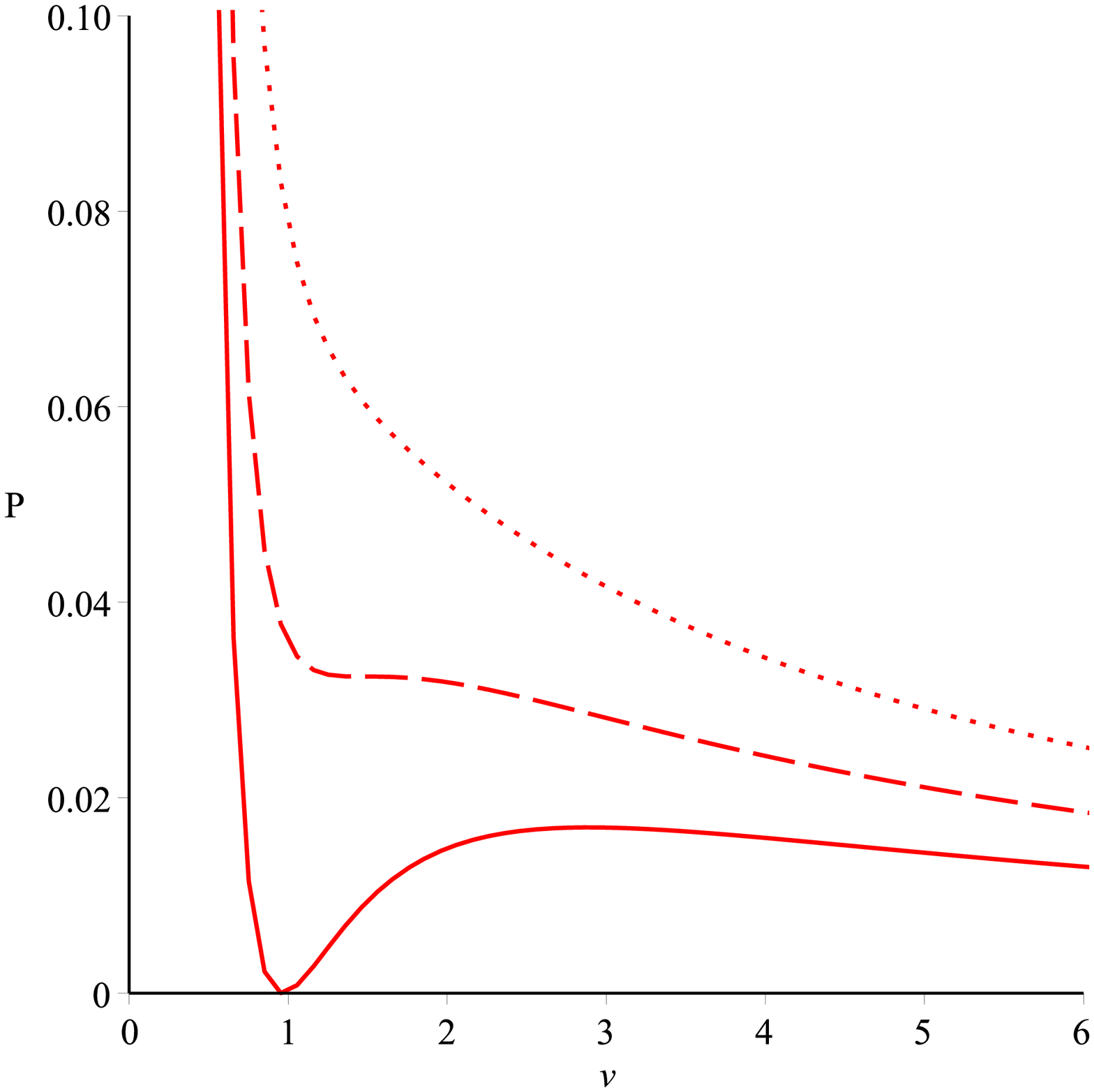}}
\caption{\label{fig:3} $P-V$ curves. (a): Solid lines correspond
to $a=0.1,\alpha=0.01$ with $T_{c}=0.397887$ and dash lines
correspond to $a=1,\alpha=0.01$ with $T_{c}=0.143605$, (b):
diagrams are plotted for $a=1, T<T_c$ with arbitrary value for
critical temperature $T_c.$ Solid red line for $\alpha=0.00001$
($T=0.8T_c$), dash blue line for $\alpha=0.01$ ($T=0.8T_c$), dash
green line for $\alpha=0.1$ ($T=0.8T_c$), dash dot orange
 line for $\alpha=0.5$ ($T=0.6T_c$) and dot black line
for $\alpha=1$ ($T=0.4T_c$). (c): Diagrams are plotted for
$a=1,\alpha=0.01$. Dot lines for upper temperature $T=0.183605$,
dash lines for critical temperature $T_{c}= 0.143605$ and solid
line for $T_{0}=0.110208.$ }
\end{figure}
By analysing the Gibbs free energy it can be revealed that at
pressure lower than the critical value the Gibbs function displays
a characteristic swallow-tail behavior which implies that at this
point the small black hole jumps into a large black hole via a
first order phase transition. It must also be noted that an
important behavior for $\alpha \to 0$ and $a \to 0$ at which
branch of small
 black hole overlaps with $T$ axis, so that small black
hole evaporates to AdS space. This means that the LBH/SBH phase
transition reduces to "BH/AdS" phase transition which is well
known as Hawking-Page phase transition. It is very interesting to
note that in the limit $\alpha \to 0$ but surrounding by a cloud
of strings, the Hawking-Page phase transition disappears and the
LBH/SBH phase transition recovers. We can interpret figure 2 in
such a way that in the absence of the GB term namely the
Schwarzschild black hole, the impact of a cloud of strings can
bring the SBH/LBH phase transition while this phase transition
does not occur in Schwarzschild type. In figure 2 we can see the
effect of GB coefficient on $G-T$ diagram which just move location
of the phase transition on the diagram.
\begin{figure}[h]
\centering \subfigure[{}]{\label{1}
\includegraphics[width=.31\textwidth]{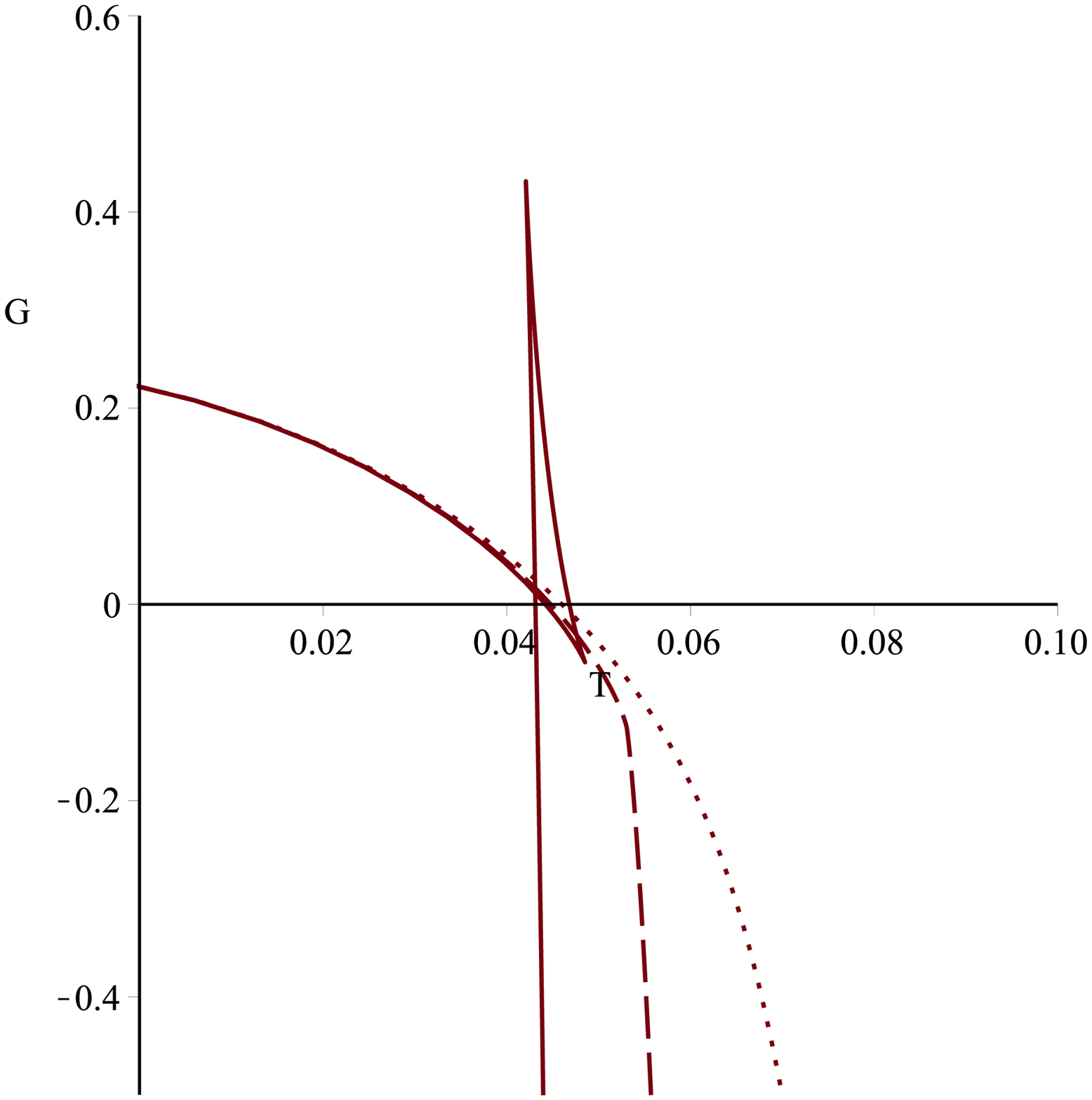}}
\hspace{1mm} \subfigure[{}]{\label{1}
\includegraphics[width=.31\textwidth]{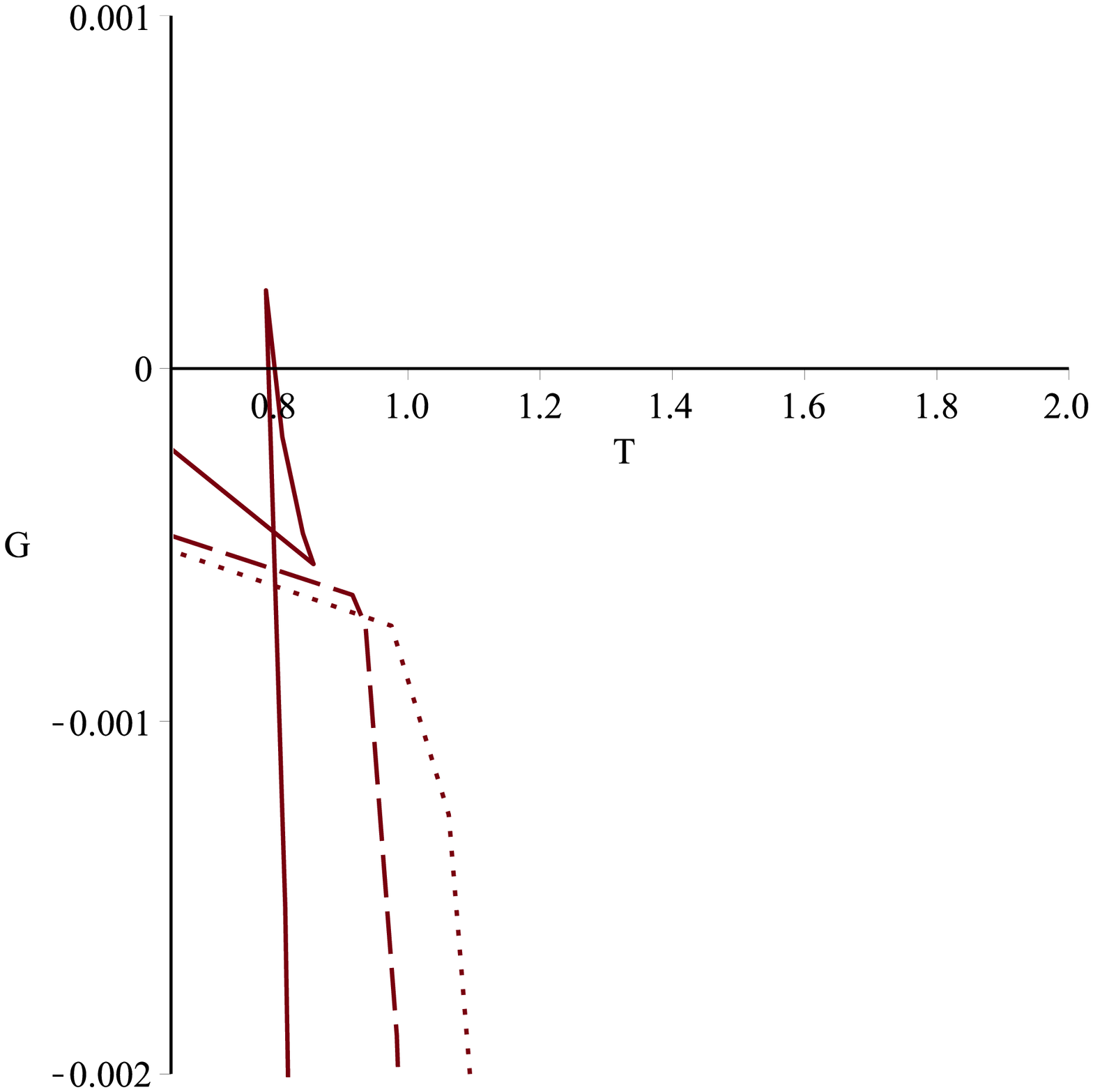}}
\subfigure[{}]{\label{1}
\includegraphics[width=.31\textwidth]{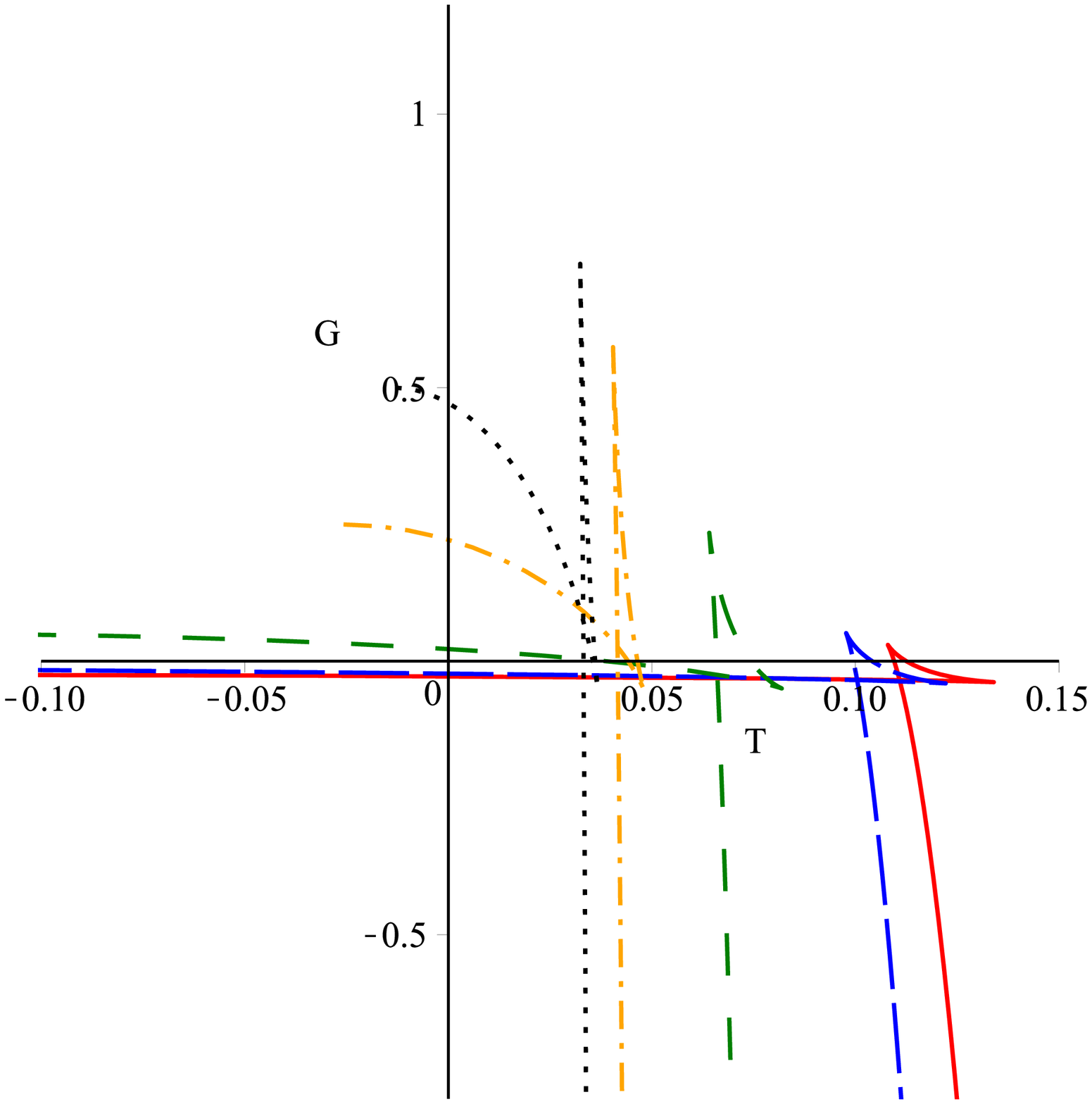}}
\hspace{1mm} \caption{$G-T$ curves. (a):  for $a=1,\alpha=0.5.$
(b): for $a=0.1,\alpha=0.001$. (c): for $a=1$ and various $\alpha$
with $P<P_c$ as indicated by solid red line for $\alpha=0.00001$
($T=0.4T_c$), dash blue line for $\alpha=0.01$ ($T=0.4T_c$), dash
green line for $\alpha=0.1$ ($T=0.4T_c$), dash dot orange line for
$\alpha=0.5$ ($T=0.5T_c$) and dot black line for $\alpha=1$
($T=0.4T_c$) and arbitrary value on $T_c.$} \label{l}
\end{figure}
The coexistence line in $P-T$ diagram in figure 3 can displays
phase transition in a different view and happens
 where two surfaces of Gibbs free energy cross each other. Since the black hole undergoes a first order phase transition so the both
  phases have the same Gibbs free energy. To study the behavior of system along the coexistence line it just enough to solve the following
  equalities between two phases at any arbitrary point on this line:
\begin{equation}
G_1=G_2,~~T_1=T_2,~~2T=T_1+T_2,
\end{equation}
which are defined by regarding to $\upsilon_1$ and $\upsilon_2$
for two different phases. The two last equalities denote to
isothermal transition. Solving these three equations we obtain
equation of pressure with respect to temperature which is plotted
in figure 3-a and 3-b for some values of $a$ parameter and
displays the effect of string clouds on $P-T$ diagram.
\begin{figure}[ht]
\centering \subfigure[{}]{\label{1}
\includegraphics[width=.35\textwidth]{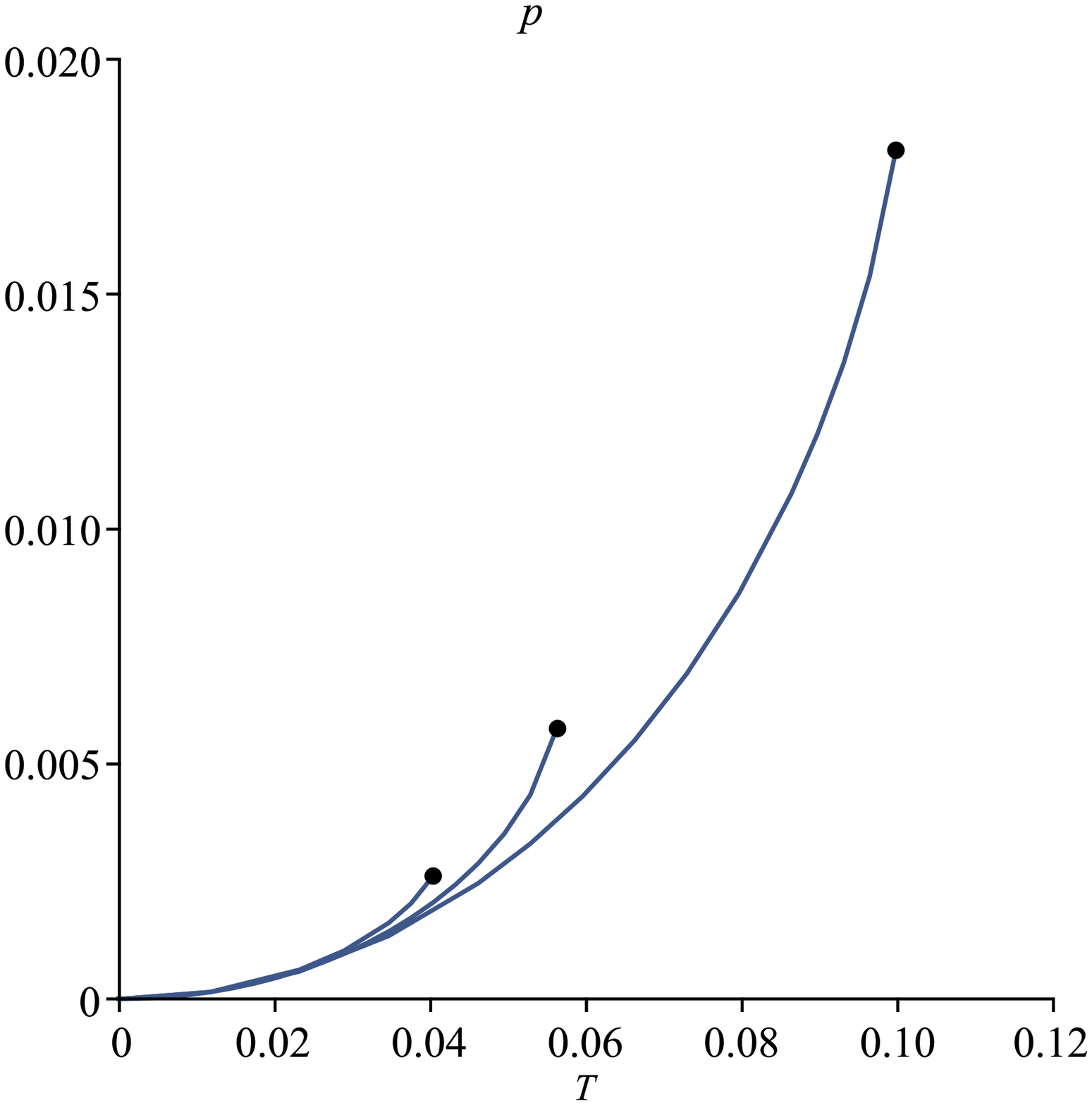}}
\hspace{1mm} \subfigure[{}]{\label{1}
\includegraphics[width=.35\textwidth]{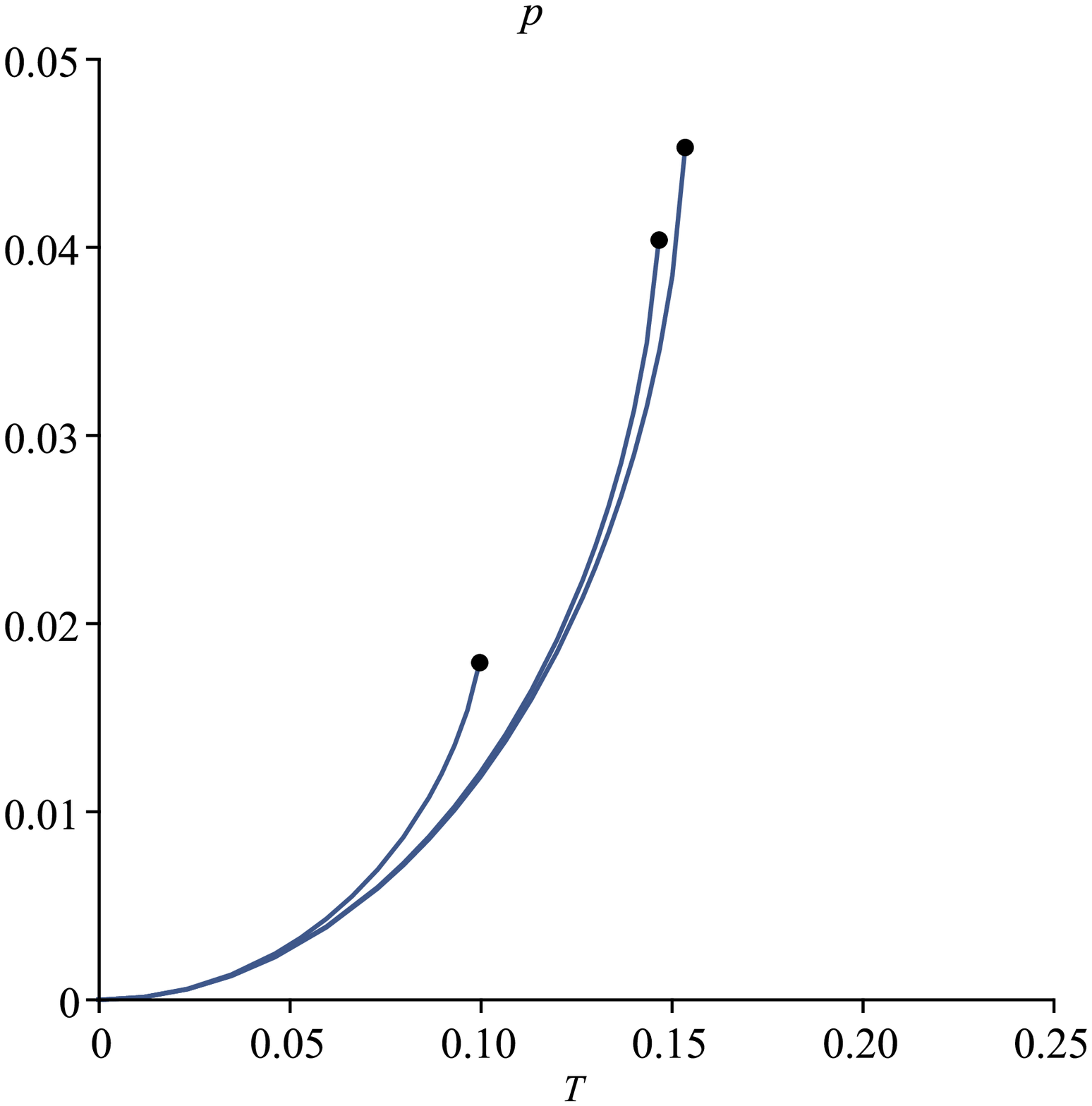}}
\caption{The coexistence lines in $p-T$ diagram which are ended at
critical points. (a): for $a=1$ and $\alpha_{GB}=\{1, 0.5, 0.1\}$
up to down, respectively. (b): for $\alpha_{GB}=1$ and $a=\{1,
0.1, 0.001\}$ up to down, respectively.} \label{l}
\end{figure}
We can also study isenthalpic curves  in $T-P$ plan and see the
behavior of our thermodynamic system in an expansion process so
called Joule-Thomson expansion [35, 36]. Originally in classical
thermodynamics this expansion describes the temperature change of
a gas or liquid through a porous plug. In this process temperature
of the system changes with respect to pressure in a constant
enthalpy and it can show heating and cooling phases. Actually sign
of the Joule-Thomson coefficient as $\mu_{JT}=\big(\frac{\partial
T}{\partial P}\big)_H$ indicates the phase of the gas. During the
expansion $\mu_{JT}>0$ denotes the cooling process in which
pressure decreases and $\mu_{JT}<0$ corresponds heating in which
pressure increases. To find out the process in our model we can
substitute $r_+$ from (2.5) into the Hawking temperature (2.7) and
plot $T-P$ curves for different constant mass $M.$ We do it in
figure 4. The process clearly has just a
 cooling phase and none of the model parameters could force it to enter a heating phase as we can see in [37] in which chemical potentials drive
  the process into a cooling-heating expansion. So in contrary with AdS black holes investigated in [37-44] our case doesn't follow a heating-cooling
   process in a Joule-Thomson expansion and there is not any inversion temperature.
\begin{figure}[ht]
\centering \subfigure[{}]{\label{1}
\includegraphics[width=.435\textwidth]{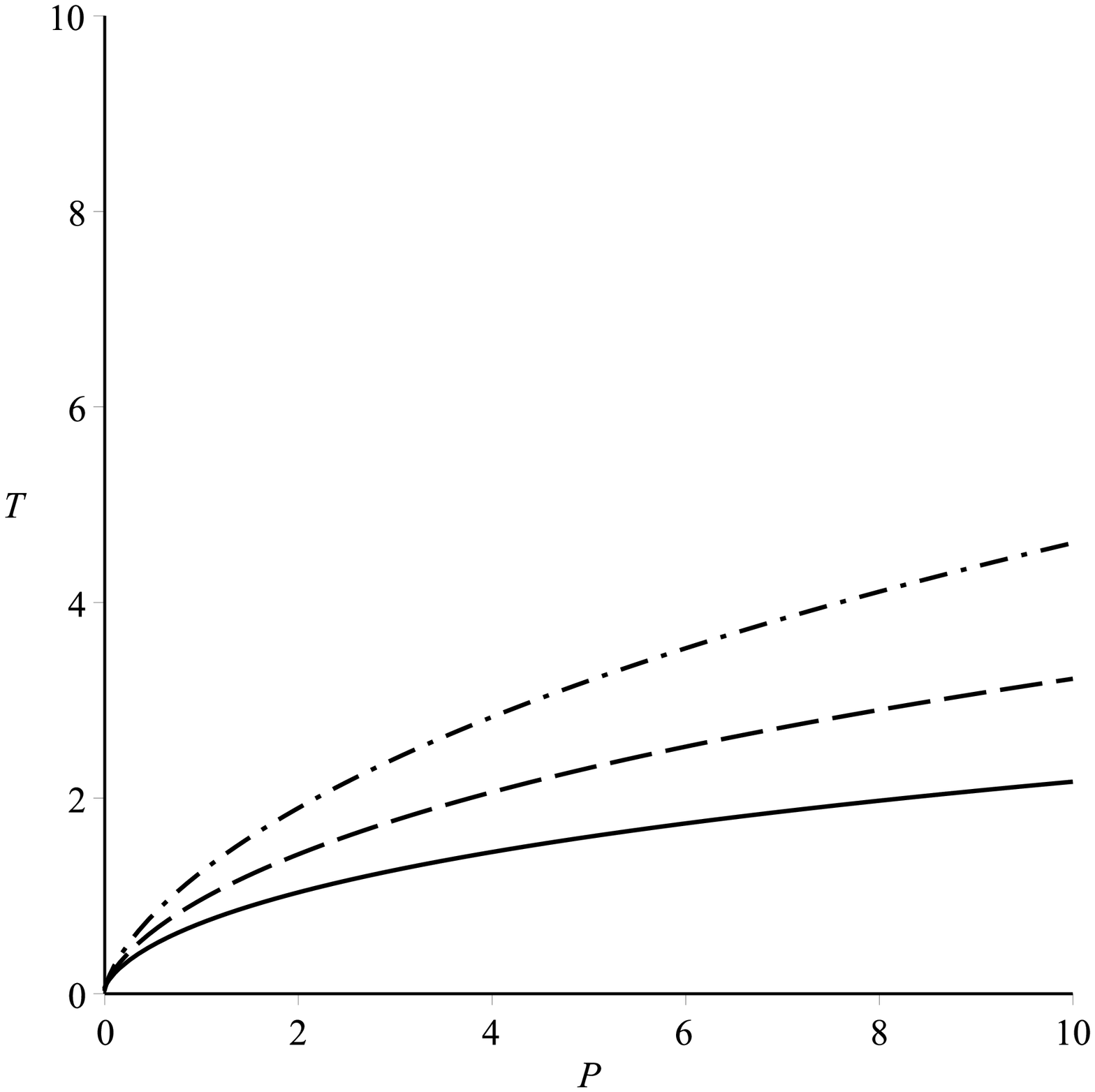}}
\hspace{1mm} \subfigure[{}]{\label{1}
\includegraphics[width=.35\textwidth]{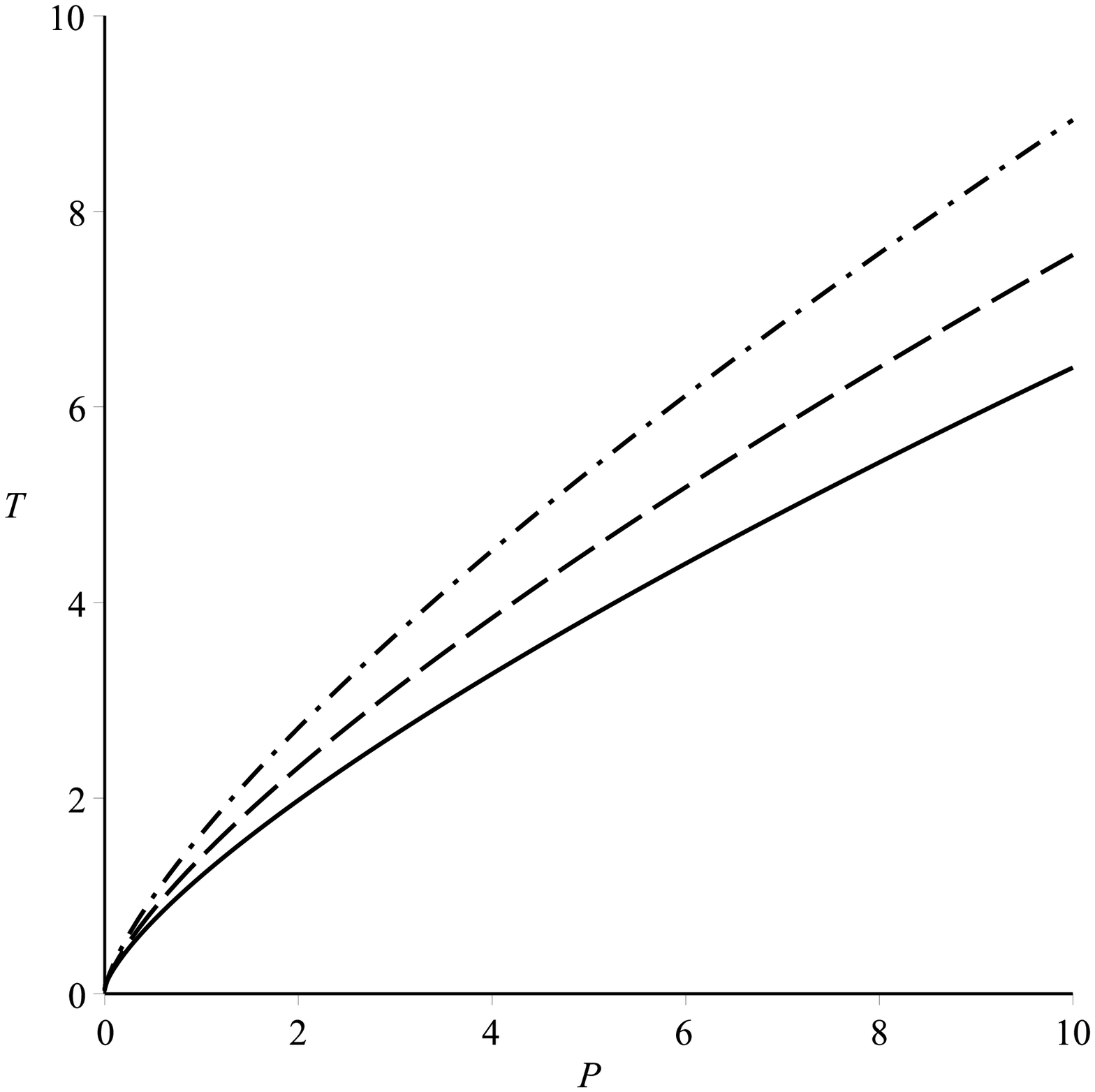}}
\caption{$T-P$ curves with isenthalpic expansion for $M=0.5$
(solid), $M=1$ (dash) $M=2$ (dash dot). (a): for $\alpha=0.1$ and
(b): $\alpha=0$} \label{l}
\end{figure}
At last, we apply the Ehrenfest's equations to study an analytic
verification of the nature of the phase transition in the critical
points. As we know from classical thermodynamics the first and
second equations of Ehrenfest are as follows [33].
\begin{eqnarray}
&&\frac{\partial P}{\partial
T}\Big|_S=\frac{C_{P2}-C_{P1}}{TV(\zeta_2-\zeta_1)}
=\frac{\Delta C_P}{TV\Delta\zeta},\label{eq:38a}\\
&&\frac{\partial P}{\partial
T}\Big|_V=\frac{\zeta_2-\zeta_1}{\kappa_{T2}-\kappa_{T1}}
=\frac{\Delta\zeta}{\Delta\kappa_{T}}\label{eq:39a}
\end{eqnarray}
where $\zeta$ and $\kappa_T$ are coefficients of the volume
expansion and isothermal compressibility of the system
respectively. The volume expansion coefficient can be written as
\begin{equation}
 V\zeta=\frac{\partial V}{\partial T}\Big|_P=\frac{\partial V}{\partial S}\Big|_P\times\frac{C_P}{T},
\end{equation}
so the first Ehrenfest's equation takes the form
\begin{equation}
\frac{\partial P}{\partial T}\Big|_{S}=\frac{\partial S}{\partial
V}\Big|_{P}.
\end{equation}
By using (3.4), the equation of (3.8) can be calculated as
$\frac{3(r_+^2+4\alpha)}{4r_+^3}=\frac{3(r_+^2+4\alpha)}{4r_+^3}$
which shows that the first equation of Ehrnfest is satisfied at
the critical point. In order to verify the second Ehrenfest's
equation, we use the thermodynamic identity $\frac{\partial
V}{\partial P}\Big|_T\times\frac{\partial P}{\partial T}\Big|_V
\times\frac{\partial T}{\partial V}\Big|_P=-1$  to obtain
isothermal compressibility coefficient as
\begin{equation}
\kappa_T =\frac{\partial T}{\partial P}\Big|_V \zeta,
\end{equation}
So the second Ehrenfest's equation would be automatically
satisfied, as well. On the other hand, the ratio of
Prigogine-Defay is calculated as follows
\begin{eqnarray}
\Pi=\frac{\Delta C_P\Delta \kappa_T}{Tv(\Delta\zeta)^2}=1.
\end{eqnarray}
So we can conclude  that the second-order phase transition occurs
in the black hole which is compatible with the transition of the
liquid-gas phase in the classical thermodynamics. Comparing our
results with the GB black holes [10], it seems quite interesting
to note that a cloud of strings does not effect on the validation
of Ehrenfest's equations.
\section{Critical exponents}
 To study the behavior of physical
quantities near the critical point it would be useful to use
rescaled quantities (3.4) for which (3.1) reaches to a re-scaled
forme as follows.
\begin{equation}
p=\bigg(\frac{T_c}{\upsilon_cP_c}\bigg)\frac{\tau}{\nu}+\bigg(\frac{64\alpha
T_c}{9\upsilon_c^3P_c}\bigg)\frac{\tau}{\nu^3}
-\bigg(\frac{2}{3\pi\upsilon_c^2P_c}\bigg)\frac{1}{\nu^2}+\bigg(\frac{8a}{27\pi\upsilon_c^3P_c}\bigg)\frac{1}{\nu^3}.
\end{equation}
The above equation of state is called as "law of corresponding
state". Defining new expansion parameters $t,\omega$ such that
\begin{gather}
  \nonumber \tau=1+t,~~~\nu=1+\omega
\end{gather}
which these are expanded around 1, we can seek thermodynamical
behavior of the system near the critical points. Therefore the law
of corresponding state (4.1) would be approximated as
\begin{equation}
p=1+At+Bt\omega+C\omega^3+\mathcal{O}(t\omega^2,\omega^4),
\end{equation}
where $C=-1$ and
\begin{eqnarray}
  A=\frac{3\sigma(a\sigma+a^2+32\alpha)}{a^3+a^2\sigma+48a\alpha+24\alpha\sigma},~~B=
  -\frac{3\sigma(a\sigma+a^2+48\alpha)}{a^3+a^2\sigma+48a\alpha+24\alpha\sigma}.
\end{eqnarray}
In the other hand critical exponents for $T<T_c$ (or $t<0$) are
defined by [32] $C_\upsilon=T\frac{\partial S}{\partial
T}\bigg|_\upsilon\propto|t|^{-\alpha},$
$\eta=\upsilon_l-\upsilon_s\propto|t|^{\beta},$ $\kappa_T=
-\frac{1}{\upsilon}\frac{\partial\upsilon}{\partial
P}\bigg|_T\propto|t|^{-\gamma},$ and
$|P-P_c|\propto|\upsilon-\upsilon_c|^{\delta}$ in which $\alpha$
shows the behavior of the specific heat at constant volume,
$\beta$ shows the isotherm behavior of the order parameter $\eta$,
$\gamma$ describes  the behavior of the isothermal compressibility
and the last exponent $\delta$ determines the behavior of pressure
in an isothermal process corresponding to $T=T_c$. The subscripts
$l$ and $s$ denote the large black
 hole and small black hole respectively in the process of phase transition.\\
As we can see from
 (2.6) the entropy is independent of $T$, so the
specific heat vanishes ($C_\upsilon=0$) and so $\alpha=0$. To
obtain the second exponent we have to evaluate $v_l$ and $v_s$ to
obtain the order parameter. The approximated pressure (4.2)
 does
not changed  during the isotherm phase transition,  which means
$p_l=p_s$ for which
\begin{equation}
Bt(\omega_l-\omega_s)+C(\omega_l^3-\omega_s^3)=0.
\end{equation}
Applying the Maxwell’s equal area law we lead to another
relationship between $\omega_l$ and $\omega_s$ such that
\begin{equation}
\int_{\omega_l}^{\omega_s}\omega\frac{dp}{d\omega}d\omega=0\rightarrow
Bt(\omega_l^2-\omega_s^2)+\frac{3}{2}C(\omega_l^3-\omega_s^3)=0.
\end{equation}
From two the above equations one can obtain
$\omega_l=-\omega_s=\sqrt{-{Bt}/{3C}}$, so
$\eta=\upsilon_l-\upsilon_s\propto\sqrt{-t}$ and therefor we can
derive $\beta=\frac{1}{2}$. The isothermal compressibility could
be calculated for $\upsilon=\upsilon_c(1+\omega)$ as follows.
\begin{equation}
\kappa_T=-\frac{1}{\upsilon_c(1+\omega)}\frac{\partial\upsilon}{\partial\omega}\frac{\partial\omega}{\partial
P}\bigg|_T
=-\frac{1}{P_c(1+\omega)}\bigg(\frac{1}{Bt}+\mathcal{O}(\omega^2)\bigg),
\end{equation}
so the third exponent is achieved as $\gamma=1$. The final
exponent is obtained in an isotherm process at critical
temperature $T=T_c$ or $t=0$,
 so from (4.2) we have $p-1=C\omega^3$ that leads to $\delta=3$.
From the results in [8] we can see that the critical exponent are
the same as the GB AdS black hole exponents, so the effect of
string cloud would not change them and both models (with and
without string cloud effects) have the same scaling laws.
\section{Conclusion}
In this paper, we extend the previous research [10] to the
extended phase space of thermodynamics and studied the critical
phenomena of the EGB black hole solution in a cloud of strings
background. Here we have considered the cosmological constant as a
 thermodynamical
 pressure. Thermodynamical quantities such as temperature, entropy, pressure and the Gibbs free energy are studied to probe the thermodynamic
  stability of the black hole. We have investigated the analogy between the EGB black holes
  surrounded by a cloud of strings and Van der Waals fluid. We successfully derived the critical points of the system and studied $P-V$ critically
  in details. It is shown that the critical quantities and Van der Waals universal ratio is affected by string cloud. It is also shown
  that in the absence of the GB term (Schwarzschild black hole), the impact of a cloud of strings can almost bring
  SBH/LBH phase transition. In the other words, in the EGB black hole with $\alpha \to 0$
   limit and surrounding by a cloud of strings, the Hawking-Page phase transition disappears and SBH/LBH phase transition recovers.
   By studying $T-P$ diagram we found this system never enter to a
   heating phase and in an Joule-Thomson expansion it cools forever. At last we investigated the behavior of our black
    hole solution near critical point by studying critical exponents and  concluded that string could not be effective and does
     not change them. As a future work it also would be interesting to study thermodynamics of the Lovelock black hole solution
      surrounded by a cloud of strings in the extended phase
      space.\\
\textbf{Acknowledgments} \\The authors are grateful to the editor
and anonymous referees for their valuable comments and suggestions
to improve the paper.
  \vskip .1cm
 \noindent
  {\bf References}\\
\begin{description}
\item[1.] S. W. Hawking, D.N. Page,~{\it {Thermodynamics of black holes in anti-de Sitter space}}, Comm. Math. Phys. {\bf 87}, 577, (1983).
\item[2.] S. Nojiri and S. D.
Odintsov "Anti de Sitter black hole thermodynamics in higher
derivative gravity and new confining
 deconfining phases in dual CFT", Phys. Lett. B 521, 87 (2001); hep-th/0109122; Erratum: Phys. Lett. B 542, 301
 (2002).
\item[3.] M. Cvetic, S. Nojiri and S. D. Odintsov, "Black hole thermodynamics and negative entropy in de Sitter and anti de Sitter Einstein-
Gauss-Bonnet gravity
 ", Nucl. Phys. B628, 295 (2022).
\item[4.] D. Kastor, S. Ray, \& J. Traschen,~{\it {  Enthalpy and mechanics of AdS black holes}}, Class. Quant. Grav. {\bf 26}, 195011, (2009).
\item[5.]  B. P. Dolan,~{\it {The cosmological constant and black hole thermodynamic potential}}, , Class. Quant. Grav. {\bf 28}, 125020, (2011).
\item[6.] D. Kubiznak \& R. B. Mann,~{\it {P-V criticality of charged AdS black holes}}, JHEP \textbf{1207}, 033, (2012).
\item[7.] S. Dutta, A. Jain and R. Soni,~{\it {Dyonic Black Hole and Holography}}, JHEP {\bf 2013}, 60, (2013); hep-th/1310.1748.
\item[8.] Gu-Qiang Li,~{\it {“Effects of dark energy on P-V criticality of charged AdS black holes}}”, Physics Letters B \textbf{735}, 256,
 (2014).
\item[9.]  M. Zhang \& W Liu,l, {\it {Coexistent physics of massive black holes in the phase transitions}}, gr-qc/1610.03648.
\item[10.]  R Cai, L. Cao, \& R. Yang,, {\it {P-V criticality in the extended phase space of Gauss-Bonnet black holes in AdS space}},
 JHEP\textbf{1309}, 005, (2013).
\item[11.]  R.  Cai, Y. Hu, Q. Pan, \& Y. Zhang, {\it { Thermodynamics of black holes in massive gravity}}, Phys. Rev. D \textbf{91}, 024032, (2015)
 hep-th/1409.2369.
\item[12.] R. Zhao, H. H. Zhao, M. S. Ma and L. C. Zhang, {\it {On the critical phenomena and thermodynamics of charged topological dilaton AdS
black holes}}, Eur. Phys. J. C \textbf{73}, 2645, (2013).
\item[13.] J. Mo, G. Li, \& X. Xu, {\it {Combined effects of f(R) gravity and conformaly invariant Maxwell field on the extended phase space
thermodynamics of higher-dimensional black holes}}, Eur. Phys.
Jour. {\bf C 76}, 545, (2016).
\item[14.] J. X. Mo, X. X. Zeng, G. Q. Li, X. Jiang, W. B. Liu, {\it {A unified phase transition picture of the charged topological black hole in
 Ho\v{r}ava-Lifshitz gravity,}}, JHEP \textbf{1310}, 056, (2013).
\item[15.] N. Altamirano, D. Kubiz\v{n}\'{a}k, R. Mann and Z. Sherkatghanad, {\it {Kerr-AdS analogue of critical point
and solid/liquid/gas phase transition}}, Class, Quantum, Grav. 31,
4, 042001, (2013), hep-th/1308.2672.
\item[16.] J. X. Mo and W. B. Liu, {\it {Ehrenfest scheme for $P$-$V$ criticality in the extended phase space of black holes}},Phys. Lett. B \textbf{727},
336, (2013).
\item[17.]  N. Altamirano, D. Kubiz\v{n}\'{a}k and R. Mann, {\it {Reentrant phase transitions in rotating AdS black holes}},Phys. Rev. D\textbf{ 88},
101502, (2013).
\item[18.]  D. C. Zou, S. J. Zhang and B. Wang, {\it { Critical behavior of Born-Infeld AdS black holes in the extended phase space thermodynamics}},
Phys. Rev. D \textbf{89}, 044002, (2014).
\item[19.]  R. Zhao, H. H. Zhao, M. S. Ma and L. C. Zhang, {\it {On the critical phenomena and thermodynamics of charged topological dilaton
AdS black holes}}, Eur. Phys. J. C \textbf{73}, 2645, (2013).
\item[20.] H. Ghaffarnejad, E. Yaraie and M. Farsam,  {\it {Quintessence Reissner Nordström anti de Sitter black holes and Joule Thomson effect.}},
Int. J. Theor. Phys, 57, 1671, (2018), gr-qc/1802.08749.
\item[21.] H. Liu and X.-h. Meng, {\it {PV criticality in the extended phase space of charged accelerating
AdS black holes}}Mod. Phys. Lett.A\textbf{31}  1650199, (2016).
\item[22.] D. Hansen, D. Kubiznak and R. B. Mann, {\it {Universality of P-V Criticality in Horizon Thermodynamics}},
 JHEP 01, 047 (2017), gr-qc/1603.05689.
\item[23.] H. Estanislao and M. G. Richarte, {\it {Black holes in Einstein–Gauss–Bonnet gravity with a string
 cloud background}}, Phys. Lett. B\textbf{ 689}, 192 (2010).
\item[24.] A. Strominger, C. Vafa, {\it {Microscopic origin of the Bekenstein-Hawking entropy}}, Phys. Lett. B \textbf{379}, 99, (1996).
\item[25.] P.~S.~Letelier, {\it {Clouds Of Strings In General Relativity}}, Phys.\ Rev.\ D {\bf 20}, 1294, (1979).
\item[26.] M.~G.~Richarte and C.~Simeone, {\it {Traversable wormholes in a string cloud}},   Int.\ J.\ Mod.\ Phys.\ D {\bf 17}, 1179,
 (2008).
\item[27.] A.~K.~Yadav, V.~K.~Yadav and L.~Yadav, {\it {Cylindrically symmetric inhomogeneous universes with a cloud of strings}},
 Int. J. Theor. Phys.  {\bf 48}, 568, (2009).
\item[28.]  A.~Ganguly, S.~G.~Ghosh and S.~D.~Maharaj, {\it {Accretion onto a black hole in a string cloud background}},
 Phys. Rev.  D {\bf 90}, 064037, (2014).
\item[29.]   K.~A.~Bronnikov, S.~W.~Kim and M.~V.~Skvortsova,, {\it {The Birkhohff theorem and string clouds}},
 Class. Quantum Grav.  {\bf 33}, 195006,
 (2016).
\item[30.]   S.~G.~Ghosh and S.~D.~Maharaj, {\it {Cloud of strings for radiating black holes in Lovelock gravity}},
 Phys. Rev. D {\bf 89}, 084027, (2014).
\item[31.]   S.~G.~Ghosh, U.~Papnoi and S.~D.~Maharaj , {\it {Cloud of strings in third order Lovelock gravity}},
Phys. Rev. D {\bf 90}, 044068, (2014).
\item[32.] T.~H.~Lee, D.~Baboolal and S.~G.~Ghosh, {\it {Lovelock black holes in a string cloud background}},
Eur. Phys. J. C {\bf 75}, 297, (2015).
\item[33.] M. W. Zemansky and R.H. Dittman, {\it {Heat and thermodynamics: an introduction}},McGraw-Hill (1997).
\item[34.] N. Goldenfeld,  "Lectures on phase transitions and the renormalization group" CRC Press, (2018).
\item[35.] D. Winterbone and A. Turan. Advanced Thermodynamics for Engineers. Butterworth-Heinemann, (1996).
\item[36.] D. C. Johnston, "Thermodynamic Properties of the van der Waals Fluid." Cond-mat.soft/1402.1205 (2014).
\item[37.] J. X. Mo and G. Q. Li, "Effects of Lovelock gravity on the Joule-Thomson expansion" gr-qc/1805.04327 (2018).
\item[38.] O. Ozgur and E. Aydner. "Joule–Thomson expansion of the charged AdS black holes", Eur. Phys. J. C 77, 24 (2017).
\item[39.] O. Ozgur and E. Aydner. "Joule–Thomson expansion of Kerr–AdS black holes," Eur. Phys. J. C 78, 123 (2018).
\item[40.] M. Chabab, H. E. Moumni, S. Iraoui, K. Masmar, and S. Zhizeh. "Joule-Thomson Expansion of RN-AdS Black Holes in $ f (R) $ gravity"
 gr-qc/1804.10042 (2018).
\item[41.] J. X. Mo, G. Q. Li, S. Q. Lan, and X. B. Xu. "Joule-Thomson expansion of $ d $-dimensional charged AdS black holes"
gr-qc/1804.02650 (2018).
\item[42.] S. Q. Lan, "Joule-Thomson expansion of charged Gauss-Bonnet black holes in AdS space." gr-qc/1805.05817 (2018).
\item[43.] C. L. A. Rizwan, A. N. Kumara, D. Vaid, and K. M. Ajith. "Joule-Thomson expansion in AdS black hole with a global monopole."
gr-qc/1805.11053 (2018).
\item[44.] Z. W. Zhao, Y. H. Xiu and N. Li. "On the throttling process of the Kerr--Newman--anti-de Sitter black holes in the extended phase
 space." gr-qc/1805.04861 (2018).
\item[45.] D. L. Wiltshire,
"Spherically symmetric solutions of Einstein-Maxwell theory with a
Gauss-Bonnet term." Phys. Lett. B 169, 36 (1986).
\item[46.] M. Brigante, H. Liu, R. C. Myers, S. Shenker, and S. Yaida. "Viscosity bound violation in higher derivative gravity."
 Phys. Rev. D 77, 126006 (2008).
\item[47.] A. Buchel and R. C. Myers. "Causality of holographic hydrodynamics." JHEP 2009, 016 (2009).
\item[48.] H. S. Reall, N. Tanahashi and B. Way. "Causality and hyperbolicity of Lovelock theories."
Class. Quantum Grav. 31, 205005 (2014).
\item[49.] R. Brustein and Y.
Sherf. "Causality violations in Lovelock theories." Phys. Rev. D
97, 084019 (2018); hep-th/1711.05140.
\item[50.] T. K. Dey, "Thermodynamics of AdS Schwarzschild black hole in the presence of external string cloud."
hep-th/1711.07008 (2017).
\item[51.] S. H. Mazharimousavi and M. Halilsoy. "Cloud of strings as source in $2+1$-dimensional $f(R)= R^{n}$ gravity."
Eur. Phys. J. C 76, 95 (2016).
\item[52.] S. G. Ghosh, S. D. Maharaj, D. Baboolal and T. H. Lee. "Lovelock black holes surrounded by quintessence.
" Eur. Phys. J. C 78, 90 (2018).
\item[53.] de M. T. Jefferson and V. B. Bezerra. "Black holes with cloud of strings and quintessence in Lovelock gravity.
" Eur. Phys. J. C 78, 534 (2018).
\item[54.] A. Anabalon, M. Appels,
R. Gregory, D. Kubiznak, R. B. Mann, and A.\"{O}vg\"{u}n.
"Holographic Thermodynamics of Accelerating Black Holes."
hep-th/1805.02687 (2018).
\item[55.] A. \"{O}vg\"{u}n, $ P-V $ criticality of a specific black hole in $ f (R) $ gravity coupled with Yang-Mills field, Adv. High Energy Phys. 8153721 (2018);
gr-qc/1710.06795.
\item[56.] R. Zhao, H.H. Zhao, M. S. Ma, and L. C. Zhang.
 "On the critical phenomena and thermodynamics of charged topological dilaton AdS
 black holes." Eur. Phys. J. C 73, 2645 (2013);gr-qc/1305.3725.
\item[57.] M. S. Ma, F. Liu, and R.Zhao. "Continuous phase transition and critical behaviors of 3D black hole with torsion." Class. Quantum  Grav.
 31, 095001 (2014); gr-qc/1403.0449.
\item[58.] I. Sakalli, and A. A.\"{O}vg\"{u}n. "Black hole radiation of massive spin-2 particles in (3+1) dimensions."
 Eur. Phys. J. Plus 131, 184 (2016); gr-qc/1605.02689.
\item[59.] I. Sakalli and A. \"{O}vg\"{u}n. "Tunnelling of vector particles from Lorentzian wormholes in 3+ 1 dimensions."
Eur. Phys. J. Plus 130, 110, (2015).
\item[60.] M. S. Ma and R. Zhao. "Phase transition and entropy spectrum of the BTZ black hole with torsion." Phys. Rev. D 89, 044005 (2014); gr-qc/1310.1491.
\item[61.] I Sakalli and A. \"{O}vg\"{u}n. "Uninformed Hawking radiation." Eur. Phys. Let. 110, 10008
(2015); gr-qc/1409.5539.
\item[62.] I. Sakalli and A. \"{O}vg\"{u}n, " Quantum tunneling of massive spin-1 particles from non-stationary metrics", Gen. Rel. Gravit. 48, 1, 1,
 (2016); gr-qc/1507.01753.
 \item[63.] H. H. Zhao, L. C. Zhang, M. S. Ma, and R. Zhao, "P-V criticality of higher dimensional charged topological dilaton
  de Sitter black holes" Phys. Rev. D90, 064018 (2014),
\end{description}
\end{document}